\documentclass[12pt]{article}
\newcounter{myfn}

\title{Normal Typicality and von Neumann's Quantum Ergodic Theorem}
\author{
Sheldon Goldstein,\footnote{Department of Mathematics, Rutgers University, 
	110 Frelinghuysen Road, Piscataway, NJ 08854-8019, USA.}
	{}\footnote{E-mail: oldstein@math.rutgers.edu}\ \ 
Joel L. Lebowitz,\setcounter{myfn}{1}\textsuperscript{\fnsymbol{myfn}}{}\footnote{E-mail: 
	lebowitz@math.rutgers.edu}\ \ 
Christian Mastrodonato,\footnote{Dipartimento di Fisica dell'Universit\`a di
	Genova and INFN sezione di Genova, Via Dodecaneso 33, 16146
	Genova, Italy.}\ \footnote{E-mail: christian.mastrodonato@ge.infn.it}\\
Roderich Tumulka,\textsuperscript{\fnsymbol{myfn}}{}\footnote{E-mail: 
	tumulka@math.rutgers.edu}\ \ and
Nino Zangh\`\i\setcounter{myfn}{4}\textsuperscript{\fnsymbol{myfn}}{}\footnote{E-mail: 
	zanghi@ge.infn.it}
}

\date{April 15, 2010}
\usepackage{amsthm,amssymb,amsmath,url,amsfonts,mathrsfs}

\theoremstyle{plain}
\newtheorem{thm}{Theorem}
\newtheorem{lem}{Lemma}

\newtheorem{defn}{Definition}

\newcommand{\ket}[1]{\vert#1\rangle}
\newcommand{\bra}[1]{\langle#1\vert}
\newcommand{\tr}{\mathrm{tr}}

\newcommand{\RRR}{\mathbb{R}}
\newcommand{\EEE}{\mathbb{E}}

\newcommand{\PPP}{\mathbb{P}}
\newcommand{\scp}[2]{\langle #1| #2 \rangle}
\newcommand{\Hilbert}{\mathscr{H}}
\newcommand{\formost}{{\textstyle\bigvee \hspace{-3.04mm} | \,\,}}
\newcommand{\be}{\begin{equation}}
\newcommand{\ee}{\end{equation}}
\renewcommand{\Re}{\mathrm{Re}}

\newcommand{\D}{D} 
\newcommand{\dd}{d} 
\newcommand{\decomp}{\mathscr{D}} 
\newcommand{\Unitary}{U} 

\newcommand{\x}[1]{{#1}}

\newcommand{\z}[1]{{#1}}

\begin{document}

\maketitle

\begin{abstract}
We discuss the content and significance of John von Neumann's quantum ergodic theorem (QET) of 1929, a strong result arising from the mere mathematical structure of quantum mechanics. The QET is a precise formulation of what we call normal typicality, i.e., the statement that, for typical large systems, every initial wave function $\psi_0$ from an energy shell is ``normal'': it evolves in such a way that $|\psi_t\rangle \langle\psi_t|$ is, for most $t$, macroscopically equivalent to the micro-canonical density matrix. The QET has been mostly forgotten after it was criticized as a dynamically vacuous statement in several papers in the 1950s. However, we point out that this criticism does not apply to the actual QET, a correct statement of which does not appear in these papers, but to a different (indeed weaker) statement. Furthermore, we formulate a stronger statement of normal typicality, based on the observation that the bound on the deviations from the average specified by von Neumann is unnecessarily coarse and a much tighter (and more relevant) bound actually follows from his proof. 

\medskip

\end{abstract}


\section{Introduction}\label{introduction}

Quantum statistical mechanics has many similarities to the classical version, and also some differences. Two facts true in the quantum but not in the classical case, \emph{canonical typicality} and (what we call) \emph{normal typicality}, follow from just the general mathematical structure of quantum mechanics. Curiously, both were discovered early on in the history of quantum mechanics, in fact both in the 1920s, and subsequently forgotten until recently. Canonical typicality was basically anticipated, though not clearly articulated, by Schr\"odinger in 1927 \cite{schr}, and rediscovered a few years ago by several groups independently \cite{GMM04,GLTZ06,PSW06}. Normal typicality, the topic of this paper, was discovered, clearly articulated, and rigorously proven by John von Neumann in 1929 \cite{vN29} as a ``quantum ergodic theorem'' (QET). In the 1950s, though, the QET was heavily criticized in two influential papers \cite{FL57,BL58} as irrelevant to quantum statistical mechanics, and indeed as dynamically vacuous. The criticisms (repeated in \cite{BL59,F61,Fbook,Lan61,Lan05}) have led many to dismiss von Neumann's QET (e.g., \cite{Lud58}, \cite[p.~273]{vH59}, \cite{PS60}, \cite{J63}, \cite{Pech84}, \cite[p.~227]{TKS91}). We show here that these criticisms are invalid. They actually apply to a statement different from (indeed weaker than) the original theorem. The dismissal of the QET is therefore unjustified. Furthermore, we also formulate two new statements about normal typicality,  see Theorem~\ref{thm:strong} and Theorem~\ref{thm:typH} below, which in fact follow from von Neumann's proof.
\z{(We provide further discussion of von Neumann's QET article in a subsequent work \cite{GLTZ10}.)}

In recent years, there has been a renewed strong interest in the foundations of quantum statistical mechanics, see \cite{GMM04,GLTZ06,PSW06, R08,RDO08,LPSW08, GLMTZ09}; von Neumann's work, which has been mostly forgotten, has much to contribute to this topic.

\bigskip

The QET concerns the long-time behavior of the quantum state vector
\be
\psi_t=\exp(-iHt)\psi_0
\ee
(where we have set $\hbar=1$) of a macroscopic quantum system, e.g., one with more than $10^{20}$ particles, enclosed in a finite volume. Suppose that $\psi_t$ belongs to a ``micro-canonical'' subspace $\Hilbert$ of the Hilbert space $\Hilbert_{\mathrm{total}}$, corresponding to an energy interval that is large on the microscopic scale, i.e., contains many eigenvalues, but small on the macroscopic scale, i.e., different energies in that interval are not discriminated macroscopically. Thus, the dimension of $\Hilbert$ is finite but huge, in fact exponential in the number of particles. We use the notation
\be
\D=\dim\Hilbert
\ee
(= $S_a$ in \cite{vN29}, $S$ in \cite{FL57,BL58}). The micro-canonical density matrix $\rho_{mc}$ is then $1/\D$ times the identity operator on $\Hilbert$, and the micro-canonical average of an observable $A$ on $\Hilbert$ is given by 
\be\label{mcav}
\tr(\rho_{mc}A)=\frac{\tr A}{\D} = \EEE \scp{\varphi}{A|\varphi}\,,
\ee
where $\varphi$ is a random vector with uniform distribution over the unit sphere of $\Hilbert$
\be
\bigl\{\varphi\in\Hilbert\,\big|\;\|\varphi\|=1\bigr\}\,,
\ee
and $\EEE$ means expectation value. In the following, we denote the time average of a function $f(t)$ by a bar,
\be
\overline{f(t)} = \lim_{T\to\infty} \frac{1}{T} \int_0^T dt \, f(t)\,.
\ee

Despite the name, the property described in the QET is not precisely analogous to the standard notion of ergodicity as known from classical mechanics and the mathematical theory of dynamical systems. That is why we prefer to call quantum systems with the relevant property ``normal'' rather than ``ergodic.'' Nevertheless, to formulate a quantum analog of ergodicity was von Neumann's motivation for the QET. It is characteristic of ergodicity that time averages coincide with phase-space averages. Put differently, letting $X_t$ denote the phase point at time $t$ of a classical Hamiltonian system, $\delta_{X_t}$ the delta measure concentrated at that point, and $\mu_{mc}$ the micro-canonical (uniform) measure on an energy surface, ergodicity is equivalent to
\be\label{ergodic}
\overline{\delta_{X_t}} =\mu_{mc}
\ee
for almost every $X_0$ on this energy surface.
In quantum mechanics, if we regard a pure state $|\psi_t\rangle \langle\psi_t|$ as analogous to the pure state $\delta_{X_t}$ and $\rho_{mc}$ as analogous to $\mu_{mc}$, the statement analogous to \eqref{ergodic} reads
\be\label{Qergodic}
\overline{\ket{\psi_t}\bra{\psi_t}} = \rho_{mc}\,.
\ee
As pointed out by von Neumann \cite{vN29}, the left hand side always exists and can be computed as follows. Let $\{\phi_\alpha\}$ be an orthonormal basis of eigenvectors of $H$ with eigenvalues $E_\alpha$.
If $\psi_0$ has coefficients $c_\alpha=\scp{\phi_\alpha}{\psi_0}$,
\be
\psi_0 = \sum_{\alpha=1}^\D c_\alpha \ket{\phi_\alpha}\,,
\ee
then
\be
\psi_t = \sum_{\alpha=1}^\D e^{-iE_\alpha t} c_\alpha \ket{\phi_\alpha}\,,
\ee
and thus
\be
\overline{\ket{\psi_t}\bra{\psi_t}}=
\sum_{\alpha,\beta} \overline{e^{-i(E_\alpha-E_\beta)t}} c_\alpha c_\beta^* \ket{\phi_\alpha}\bra{\phi_\beta}\,.
\ee
Suppose that $H$ is non-degenerate; then $E_\alpha-E_\beta$ vanishes only for $\alpha=\beta$, so the time averaged exponential is $\delta_{\alpha\beta}$, and we have that
\be
\overline{\ket{\psi_t}\bra{\psi_t}}
=\sum_{\alpha} |c_\alpha|^2 \ket{\phi_\alpha}\bra{\phi_\alpha}\,.
\ee
While the case \eqref{Qergodic} occurs only for those special wave functions that have $|c_\alpha|^2=1/\D$ for all $\alpha$, in many cases it is true of \emph{all} initial wave functions $\psi_0$ on the unit sphere of $\Hilbert$ that $\overline{\ket{\psi_t}\bra{\psi_t}}$ is \emph{macroscopically equivalent} to $\rho_{mc}$.

\z{What we mean here by macroscopic equivalence corresponds in the work of von Neumann \cite{vN29} 
to a decomposition of $\Hilbert$ into mutually orthogonal subspaces
$\Hilbert_\nu$, \be\label{decomp} \Hilbert =
\bigoplus_\nu \Hilbert_\nu\,, \ee such that each 
$\Hilbert_\nu$ corresponds to a different macro-state $\nu$. We call the $\Hilbert_\nu$ the ``macro-spaces'' and} write $\decomp$ for
the family $\{\Hilbert_\nu\}$ of subspaces, called a ``macro-observer'' in
von Neumann's paper, and $P_\nu$ for the projection to $\Hilbert_\nu$. We
use the notation \be\label{dnudef} \dd_\nu =\dim \Hilbert_\nu \ee (=
$s_{\nu,a}$ in \cite{vN29}, $s_\nu$ in \cite{FL57,BL58}).\footnote{Von
Neumann motivated the decomposition \eqref{decomp} by beginning with a
family of operators corresponding to coarse-grained macroscopic observables
and arguing that by ``rounding'' the operators, the family can be converted
to a family of operators $M_1,\ldots,M_k$ that commute with each other,
have pure point spectrum, and have huge degrees of degeneracy. (This
reasoning has inspired research about whether for given operators
$A_1,\ldots,A_k$ whose commutators are small one can find approximations
$M_i\approx A_i$ that commute exactly; the answer is, for $k\geq 3$ and
general $A_1,\ldots,A_k$, no \cite{Choi88}.)
A macro-state can then be characterized by a list $\nu=(m_1,\ldots,m_k)$ of
eigenvalues $m_i$ of the $M_i$, and corresponds to the subspace
$\Hilbert_\nu \subseteq \Hilbert$ containing the simultaneous eigenvectors
of the $M_i$ with eigenvalues $m_i$; that is, $\Hilbert_\nu$ is the
intersection of the respective eigenspaces of the $M_i$ and $\dd_\nu$ is
the degree of simultaneous degeneracy of the eigenvalues
$m_1,\ldots,m_k$. For a notion of macro-spaces that does not require that
  the corresponding macro-observables commute, see \cite{DRMN06}, in
  particular Section 2.1.1. (Concerning the main results discussed below,
  Theorems 1 and 2, a plausible guess is that normal typicality extends
  to non-commuting families $A_1,\ldots,A_k$---of observables that may also
  fail to commute with $\rho_{mc}$--- provided that the observables have a
  sufficiently small variance in the sense of Lemma 1 below, i.e., that $
  Var\left( \langle\varphi| A| \varphi\rangle\right)$ be small. We
  shall however not elaborate on this here.)}


As a simple example, we may consider, for a gas consisting of $n>10^{20}$
atoms enclosed in a box $\Lambda\subset \RRR^3$, the following 51
macro-spaces $\Hilbert_0,\Hilbert_2,\Hilbert_4,\ldots,\Hilbert_{100}$:
$\Hilbert_\nu$ contains the quantum states for which the number of atoms in
the left half of $\Lambda$ lies between $\nu-1$ percent of $n$ and $\nu+1$
percent of $n$. Note  that in this example
$\Hilbert_{50}$ has the overwhelming majority of dimensions.\footnote{Actually, these subspaces form an orthogonal
decomposition of $\Hilbert_{\mathrm{total}}$ rather than of the energy
shell $\Hilbert$, since the operator of particle number in the left half of
$\Lambda$ fails to map $\Hilbert$ to itself. Thus, certain approximations
that we do not want to describe here are necessary in order to obtain an
orthogonal decomposition of $\Hilbert$.}

Given $\decomp$, we say that two density matrices $\rho$ and $\rho'$ are \emph{macroscopically equivalent}, in symbols
\be\label{macroequiv}
\rho \stackrel{\decomp}{\sim} \rho'\,,
\ee
if and only if
\be
\tr(\rho P_\nu) \approx \tr(\rho' P_\nu)
\ee
for all $\nu$. (The sense of $\approx$ will be made precise later.) For example, $\ket{\psi}\bra{\psi}\stackrel{\decomp}{\sim}\rho_{mc}$ if and only if
\be\label{approx1}
\|P_\nu \psi\|^2 \approx \frac{\dd_\nu}{\D}
\ee
for all $\nu$. This is, in fact, the case for most vectors $\psi$ on the
unit sphere of $\Hilbert$, provided the $\dd_\nu$ are sufficiently
large, as follows, see \eqref{cheb}, from the following easy geometrical fact, see e.g., \cite[p.~55]{vN29}; see also Appendix II of \cite{J63}. 

\begin{lem}\label{lem:EEE}
If $\Hilbert_\nu$ is any fixed subspace of dimension $\dd_\nu$ and $\varphi$ is a random vector with uniform distribution on the unit sphere then
\be\label{EEE}
\EEE \|P_\nu \varphi\|^2 = \frac{\dd_\nu}{\D}\,,\quad
Var \|P_\nu\varphi\|^2 = \EEE\Bigl(\|P_\nu\varphi\|^2-\frac{\dd_\nu}{\D}\Bigr)^2=
\frac1{d_\nu}\Bigl(\frac{d_\nu}{D}\Bigr)^2 \frac{(D- d_\nu)}{( D+1)}\,.
\ee
\end{lem}

Returning to the time average, we obtain that $\overline{\ket{\psi_t}\bra{\psi_t}}\stackrel{\decomp}{\sim}\rho_{mc}$ if and only if
\be\label{mcappear}
\sum_\alpha |c_\alpha|^2 \scp{\phi_\alpha}{P_\nu|\phi_\alpha} \approx \frac{\dd_\nu}{\D}
\ee
for all $\nu$. Condition \eqref{mcappear} is satisfied for every $\psi_0\in\Hilbert$ with $\|\psi_0\|=1$ if
\be
\scp{\phi_\alpha}{P_\nu|\phi_\alpha} \approx \frac{\dd_\nu}{\D}
\ee
for every $\alpha$ and $\nu$, a condition on $H$ and $\decomp$
that von Neumann showed is typically obeyed, in a sense which we
shall explain. The analogy between
$\overline{\ket{\psi_t}\bra{\psi_t}}\stackrel{\decomp}{\sim}\rho_{mc}$ and
ergodicity lies in the fact that the time average of a pure state in a
sense agrees with the micro-canonical ensemble, with the two differences
that the agreement is only an approximate agreement on the macroscopic
level, and that it typically holds for \emph{every}, rather than \emph{almost every}, pure state.

However, even more is true for many quantum systems: Not just the time average but even $\ket{\psi_t}\bra{\psi_t}$ itself is macroscopically equivalent to $\rho_{mc}$ for most times $t$ in the long run, i.e.,
\be\label{approx}
\|P_\nu \psi_t\|^2 \approx \frac{\dd_\nu}{\D}
\ee
for all $\nu$ for most $t$. Such a system, defined by $H$, $\decomp$, and $\psi_0$, we call \emph{normal}, a terminology inspired by the concept of a \emph{normal real number} \cite{normalnumber}.
Above we have stressed the continuity with the standard notion of  
ergodicity. Yet, normality is in part stronger than ergodicity (it involves no time-averaging)   
and in part weaker (it involves only macroscopic equivalence); in short, it is a different notion.

\medskip

Suppose now, \z{as in the example between \eqref{dnudef} and \eqref{macroequiv},} that one of the macro-spaces, $\Hilbert_\nu=\Hilbert_{eq}$, has the overwhelming majority of dimensions,
\be\label{eq}
\frac{d_{eq}}{D} \approx 1\,.
\ee
It is then appropriate to call this macro-state the thermal equilibrium state and write $\nu=eq$. We say that \emph{the system is in thermal equilibrium} at time $t$ if and only if
$\|P_{eq}\psi_t\|^2$ is close to 1, or, put differently, if and only if
\be\label{eqdef}
\|P_{eq}\psi_t\|^2\approx
\frac{\dd_{eq}}{\D}\,.
\ee
Thus, if a system is normal then it is in equilibrium most of the time. Of course, if it is not in equilibrium initially, the waiting time until it first reaches equilibrium is not specified, and may be longer than the present age of the universe.\footnote{Furthermore, due to the quasi-periodicity of the time-dependence of any density matrix (not just a pure one) of our system, it will keep keep on returning to (near) its initial state.} 

The case that one of the $\Hilbert_\nu$ has the overwhelming majority of dimensions is an important special case but was actually not considered by von Neumann; it is discussed in detail in \cite{GLMTZ09}. Von Neumann (and many other authors) had a different understanding of thermal equilibrium; he would have said a system is in thermal equilibrium at time $t$ if and only if \eqref{approx} holds for all $\nu$, so that $\ket{\psi_t}\bra{\psi_t}\stackrel{\decomp}{\sim}\rho_{mc}$. Here we disagree with him, as well as with his suggestion that the further theorem in \cite{vN29}, which he called the ``quantum $H$-theorem'' and which is a close cousin of the QET, is a quantum analog of Boltzmann's $H$-theorem.  Yet other definitions of thermal equilibrium have been used in \cite{R08,LPSW08}; see Section 6 of \cite{GLMTZ09} for a comparative overview, \z{and \cite{GLTZ10} for a broader overview of such definitions.}

\medskip

The QET provides conditions under which a system is normal for \emph{every} initial state vector $\psi_0$. Note that statements about \emph{most} initial state vectors $\psi_0$ are much weaker; for example, \emph{most} state vectors $\psi_0$ are in thermal equilibrium by Lemma~\ref{lem:EEE}, so a statement about \emph{most} $\psi_0$ need not convey any information about systems starting out in non-equilibrium. Furthermore, the QET asserts \emph{normal typicality}, i.e., that typical macroscopic systems are normal for every $\psi_0$; more precisely, that for \emph{most} choices of $\decomp$ (or $H$), macroscopic systems are normal for every $\psi_0$. It thus provides reason to believe that macroscopic systems in practice are normal. 

{\em  Informal statement of the QET (for fully precise statements  
see Theorems~\ref{thm:vN}--\ref{thm:typH} below):} Following von  
Neumann, we say that a Hamiltonian $H$ with non-degenerate  
eigenvalues $E_1,\ldots,E_\D$ \emph{has no resonances} if and only if
\be\label{noresonance}
E_{\alpha}-E_{\beta} \neq E_{\alpha'}-E_{\beta'}
\text{ unless }\begin{cases}\text{either } \alpha= \alpha' \text 
{ and } \beta= \beta' \\
\text{or }\alpha=\beta \text{ and }\alpha'=\beta'\,.\end{cases}
\ee
In words, this means that also the energy differences are non- 
degenerate. \x{Let $\Hilbert$ be any Hilbert space} of finite dimension $ 
\D$, and let $H$ be a self-adjoint operator on $\Hilbert$ with no  
degeneracies and no resonances. If the natural numbers $\dd_\nu$ are  
sufficiently large (precise conditions will be given later) and $\sum_ 
\nu \dd_\nu =\D$, then most families $\decomp = \{\Hilbert_\nu\}$ of  
mutually orthogonal subspaces $\Hilbert_\nu$ with $\dim\Hilbert_\nu= 
\dd_\nu$ are such that for every wave function $\psi_0\in\Hilbert$  
with $\|\psi_0\|=1$ and every $\nu$, \eqref{approx} holds most of the  
time in the long run.

\bigskip

When we say that a statement $p(x)$ is true ``for most $x$'' we mean that
\be\label{formostdef2}
\mu\{x|p(x)\} \geq 1-\delta\,,
\ee
where $0<\delta \ll 1$, and $\mu$ is a suitable probability measure; we will always use the appropriate \emph{uniform} measure, as specified explicitly in Section~\ref{sec:most}. (When we speak of ``most of the time in the long run'', the meaning is a bit more involved since there is no uniform probability measure on the half axis $[0,\infty)$; see Section~\ref{sec:most}.)

Let $p(\decomp,\psi_0)$ be the statement that for every $\nu$, \eqref{approx} holds most of the time in the long run. The misunderstanding of the QET starting in the 1950s consists of mixing up the statement
\be\label{mostDallpsip}
\text{for most }\decomp:\: \text{for all }\psi_0:\: p(\decomp,\psi_0)\,,
\ee
which is part of the QET, with the inequivalent statement 
\be\label{allpsimostDp}
\text{for all }\psi_0:\: \text{for most }\decomp:\: p(\decomp,\psi_0)\,.
\ee
To see that these two statements are indeed inequivalent, let us illustrate the difference between ``for most $x$: for all $y$: $p(x,y)$'' and ``for all $y$: for most $x$: $p(x,y)$'' by two statements about a company:
\be\label{Sa}
\mbox{
\begin{minipage}{0.5\textwidth}
\textit{Most employees are never ill.}
\end{minipage}}
\ee
\be\label{Sb}
\mbox{
\begin{minipage}{0.5\textwidth}
\textit{On each day, most employees are not ill.}
\end{minipage}}
\ee
Here, $x$ ranges over employees, $y$ over days, and $p(x,y)$ is the statement ``Employee $x$ is not ill on day $y$.'' It is easy to understand that \eqref{Sa} implies \eqref{Sb}, and \eqref{Sb} does not imply \eqref{Sa}, as there is the (very plausible) possibility that most employees are sometimes ill, but not on the same day.

Von Neumann's proof establishes \eqref{mostDallpsip}, while the proofs in \cite{FL57,BL58} establish only the weaker version \eqref{allpsimostDp}. Von Neumann also made clear in a footnote on p.~58 of his article \cite{vN29} which version he intended:
\begin{quotation}
Note that what we have shown is not that for every given $\psi$ or $A$ the ergodic theorem and the $H$-theorem hold for most $\omega_{\lambda,\nu,a}$, but instead that they hold universally for most $\omega_{\lambda,\nu,a}$, i.e., for all $\psi$ and $A$. The latter is of course much more than the former.
\end{quotation}
Here, $A$ is not important right now while $\omega_{\lambda,\nu,a}$ corresponds to $\decomp$ in our notation. So the quotation means that what von Neumann has shown is not \eqref{allpsimostDp} but \eqref{mostDallpsip} for a certain $p$.

\bigskip

The remainder of this paper is organized as follows. In Section~\ref{sec:most} we make explicit which measures are used in the role of $\mu$. In Section~\ref{sec:bounds} we give the precise definition of normality. Section~\ref{sec:vN} contains a precise formulation of von Neumann's theorem and an outline of his proof. Section~\ref{sec:strong} contains our stronger version of the QET with tighter bounds on the deviations. In Section~\ref{sec:overview} we show that the versions of the QET in \cite{FL57,BL58} differ from the original as described above. In Section~\ref{sec:typicalH}, we provide another version of the QET, assuming typical $H$ instead of typical $\decomp$. Finally, in Section~\ref{sec:history} we compare von Neumann's result with  recent literature.

\section{Measures of ``Most''}
\label{sec:most}

Let us specify which measure $\mu$ is intended in \eqref{formostdef2} when referring to most wave functions, most unitary matrices, most orthonormal bases, most Hamiltonians, most subspaces, or most decompositions $\decomp$. It is always the appropriate uniform probability measure. 

For wave functions $\psi$, $\mu$ is the (normalized, $(2\D-1)$-dimensional) surface area measure on the unit sphere in Hilbert space $\Hilbert$.

For unitary matrices $U=(U_{\alpha\beta})$, the uniform probability distribution over the unitary group $\Unitary(\D)$ is known as the \emph{Haar measure}. It is the unique normalized measure that is invariant under multiplication (either from the left or from the right) by any fixed unitary matrix. 

For orthonormal bases, the Haar measure defines a probability distribution (the \emph{uniform distribution}) over the set  of \x{orthonormal bases of $\Hilbert$, $ONB(\Hilbert)$, as} follows. Fix first some orthonormal basis $\phi_1,\ldots,\phi_\D$ for reference. Any other orthonormal basis $\omega_1,\ldots,\omega_\D$ can be expanded into the $\phi_\beta$,
\be\label{omegaUphi}
\omega_\alpha = \sum_{\beta=1}^\D U_{\alpha\beta} \phi_\beta\,,
\ee
where the coefficients $U_{\alpha\beta}$ form a unitary matrix. Conversely, for any given unitary matrix $U=(U_{\alpha\beta})$, \eqref{omegaUphi} defines an orthonormal basis; thus, a random Haar-distributed $U$ defines a random orthonormal basis $(\omega_\alpha)$, whose distribution we call the uniform distribution. It is independent of the choice of the reference basis $\phi$ because the Haar measure is invariant under right multiplication by a fixed unitary matrix. Note also that the marginal distribution of any single basis vector $\omega_\alpha$ is the uniform distribution on the unit sphere in $\Hilbert$.

For Hamiltonians, we will regard the eigenvalues as fixed and consider the uniform measure for its eigenbasis. This is the same distribution as that of $H=UH_0U^{-1}$ when $U$ has uniform distribution and $H_0$ is fixed.

For subspaces, we will regard the dimension $\dd$ as fixed; the measure over all subspaces of dimension $\dd$ arises from the measure on $ONB(\Hilbert)$ as follows. If the random orthonormal basis $\omega_1,\ldots,\omega_\D$ has uniform distribution, we consider the random subspace spanned by $\omega_1,\ldots,\omega_\dd$ and call its distribution uniform.

For decompositions $\decomp=\{\Hilbert_\nu\}$, we will regard the number $N$ of subspaces as fixed, as well as their dimensions $\dd_\nu$; the measure over decompositions arises from the measure on $ONB(\Hilbert)$ as follows. Given the orthonormal basis $\omega_1,\ldots,\omega_\D$, we let $\Hilbert_\nu$ be the subspace spanned by those $\omega_\alpha$ with $\alpha\in J_\nu$, where the index sets $J_\nu$ form a partition of $\{1,\ldots,\D\}$ with $\#J_\nu =\dd_\nu$; we also regard the index sets $J_\nu$ as fixed.

The Haar measure is also invariant under the inversion $U\mapsto U^{-1}$. A consequence is what we will call the ``unitary inversion trick'': If $\phi$ is any fixed orthonormal basis and $\omega$ a random orthonormal basis with uniform distribution then the joint distribution of the coefficients $U_{\alpha\beta}=\scp{\phi_\beta}{\omega_\alpha}$ is the same as if $\omega$ were any fixed orthonormal basis and $\phi$ random with uniform distribution. The reason is that in the former case 
the matrix $U$ is Haar-distributed, and in the latter case 
$U^{-1}$ is Haar-distributed,  which yields the same distribution of $U$. As a special case, considering only one of the $\omega_\alpha$ and calling it $\psi$, we obtain that if $\phi$ is any fixed orthonormal basis and $\psi$ a random vector with uniform distribution then the joint distribution of the coefficients $\scp{\phi_\beta}{\psi}$ is the same as if $\psi$ were any fixed unit vector and $\phi$ random with uniform distribution. 

The concept of ``most times'' is a little more involved because it involves a limiting procedure. Let $\delta'>0$ be given; we say that a statement $p(t)$ \emph{holds for $(1-\delta')$-most $t$} (in the long run) if and only if
\be\label{mostt}
\liminf_{T\to\infty} \frac{1}{T} \biggl|\Bigl\{0<t<T\Big| p(t) \text{ holds}
\Bigr\}\biggr| \geq 1-\delta'\,,
\ee
where $|M|$ denotes the size (Lebesgue measure) of the set $M\subseteq \RRR$. (So this concept of ``most'' does not directly correspond to a probability distribution.)

\section{The Method of  Appeal to Typicality}

We would like to clarify the status  of statements about ``most'' $\decomp$ (or, for that matter, most $H$ or most $\psi_0$), and in so doing elaborate on von Neumann's method of appeal to typicality. In 1955, Fierz   criticized  this method as follows 
 \cite[p.~711]{F55}:\footnote{This quotation was translated from the German by R.~Tumulka.}
\begin{quotation}
The physical justification of the hypothesis [that all observers are equally probable] is of course questionable, as the assumption of equal probability for all observers is entirely without reason. Not every macroscopic observable in the sense of von Neumann will really be measurable. Moreover, the observer will try to measure exactly those quantities which appear characteristic of a given system. 
\end{quotation}
In the same vein, Pauli  wrote 
in a private letter to Fierz in 1956 \cite{P56}:
\begin{quotation}
As far as assumption B [that all observers are equally probable] is concerned [\ldots] I consider it \emph{now} not only as lacking in plausibility, but \emph{nonsense}. 
\end{quotation}
Concerning these objections, we first note that it is surely
informative that normality holds for some $\decomp$s, let alone that it
holds in fact for most $\decomp$s, with ``most'' understood in a
mathematically natural way. But we believe that more should be said. 

When  employing the method of appeal to typicality, one usually uses 
 the language of probability theory. When we do so  we do not mean to imply that any of the objects considered is random in reality. What we mean is that certain sets (of wave functions, of orthonormal bases, etc.)\ have certain sizes (e.g., close to 1) in terms of certain natural measures of size. That is, we describe the behavior that is \emph{typical} of wave functions, orthonormal bases, etc.. However, since the mathematics is equivalent to that of probability theory, it is convenient to adopt that language. For this reason, we do not mean, when using a normalized measure $\mu$, to make an ``assumption of a priori probabilities,'' even if we use the word ``probability.'' Rather, we have in mind that, if a condition is true of most $\decomp$, or most $H$, this fact may {\em suggest} that the condition is also true of a concrete given system, unless we have reasons to expect otherwise.

Of course, a theorem saying that a condition is true of the vast majority of systems does not \emph{prove} anything about a concrete given system; if we want to know for sure whether a given system is normal for every initial wave function, we need to check the relevant condition, which is \eqref{cond2} below. Nevertheless, a typicality theorem is, as we have suggested, illuminating; at the very least, it is certainly useful to know which behaviour is typical and which is exceptional. Note also that the terminology of calling a system ``typical'' or ``atypical'' might easily lead us to wrongly conclude that an  atypical system  will not be normal. A given system may have some properties that are atypical and nevertheless satisfy the condition \eqref{cond2} implying that the system is normal for every initial wave function.

The method of appeal to typicality belongs to a long tradition in physics, which includes also Wigner's work on random matrices of the 50s. In the words of  
Wigner \cite{Wigner}:
\begin{quote}
One [\dots] deals with a specific system, 
with its proper (though in many cases unknown) Hamiltonian, yet pretends 
that one deals with a multitude of systems, all with their own Hamiltonians, 
and averages over the properties of these systems. Evidently, such a procedure 
can be meaningful only if it turns out that the properties in which one is interested are the same for the vast majority of the admissible Hamiltonians.
\end{quote}
This method was used by Wigner to obtain specific new and surprising predictions about detailed properties of complex quantum systems in nuclear physics. Here the method of appeal to typicality is used to establish much less, viz., approach to thermal equilibrium.

\section{Bounds on Deviations}
\label{sec:bounds}

Two different definitions of normality are relevant to our discussion. Consider a system for which $\Hilbert, H, \decomp$, and $\psi_0$ are given. Let $N$ denote the number of macro-spaces $\Hilbert_\nu$, and let $\varepsilon>0$ and $\delta'>0$ also be given.

\begin{defn}
The system is $\varepsilon$-$\delta'$-normal in von Neumann's \cite{vN29} sense if and only if, for $(1-\delta')$-most $t$ in the long run,
\be\label{vNdef}
\Bigl|\|P_\nu\psi_t\|^2 - \frac{\dd_\nu}{\D} \Bigr|< \varepsilon \sqrt{\frac{\dd_\nu}{N\D}}
\ee
for all $\nu$.\footnote{Let us connect this to how von Neumann formulated the property considered in the QET, which is: for $(1-\delta')$-most $t$ in the long run,
\be\label{vNdeforig}
\bigl|\scp{\psi_t}{A|\psi_t} - \tr \, A/\D \bigr|< \varepsilon \sqrt{\tr(A^2)/\D}
\ee
for every real-linear combination (``macro-observable'') $A=\sum_\nu \alpha_\nu P_\nu$. The quantity $\tr\, A/\D=\tr(\rho_{mc}A)$ is the micro-canonical average of the observable $A$. The quantity $\sqrt{\tr(A^2)/\D}=\sqrt{\tr(\rho_{mc} A^2)}$ was suggested by von Neumann as a measure of the magnitude of the observable $A$ in the micro-canonical average. To see that \eqref{vNdeforig} is more or less equivalent to \eqref{vNdef}, note first that \eqref{vNdeforig} implies, by setting one $\alpha_\nu=1$ and all others to zero, that
\be\label{vNerror}
\bigl|\|P_\nu\psi_t\|^2 - \dd_\nu/\D \bigr|< \varepsilon \sqrt{\dd_\nu/\D}\,.
\ee
This is only slightly weaker than \eqref{vNdef}, namely by a factor of $\sqrt{N}$, when $N$ is much smaller than $\D/\dd_\nu$, as would be the case for the $\Hilbert_\nu$ considered by von Neumann. Conversely, \eqref{vNdef} for every $\nu$ implies \eqref{vNdeforig} for every $A$: This follows from
\be
\sum_\nu |x_\nu| \leq \sqrt{N} \sqrt{\sum_\nu |x_\nu|^2}\,,
\ee
a consequence of the Cauchy--Schwarz inequality, by setting $x_\nu = \alpha_\nu\varepsilon \sqrt{\dd_\nu/N\D}$.}
\end{defn}

\begin{defn}
The system is $\varepsilon$-$\delta'$-normal in the strong sense if and only if, for $(1-\delta')$-most $t$ in the long run,
\be\label{strongdef}
\Bigl|\|P_\nu\psi_t\|^2 - \frac{\dd_\nu}{\D} \Bigr|< \varepsilon \frac{\dd_\nu}{\D}
\ee
for all $\nu$.
\end{defn}

In the cases considered by von Neumann (\ref{strongdef}) is a much stronger inequality than (\ref{vNdef}).
The motivation for considering \eqref{strongdef} is twofold. On the one hand, Lemma~\ref{lem:EEE} implies that for most wave functions $\varphi$, the deviation of $\|P_\nu\varphi\|^2$ from $\dd_\nu/\D$ is actually smaller than $\dd_\nu/\D$. (Indeed, the Chebyshev inequality yields for $X=\|P_\nu\varphi\|^2$ that
\be\label{cheb}
\mu\Bigl( |X-\dd_\nu/\D| < \varepsilon \frac{\dd_\nu}{\D} \Bigr) \geq 1-\frac{Var X}{(\varepsilon \dd_\nu/\D)^2} \geq 1- \frac{1}{\varepsilon^2 \dd_\nu}\,,
\ee
which tends to 1 as $\dd_\nu\to\infty$.) On the other hand, strong normality means that $\|P_\nu\psi_t\|^2$ actually is \emph{close} to $\dd_\nu/\D$, as the \emph{relative error} is small. In contrast, the bound in \eqref{vNdef} is greater than the value to be approximated, and so would not justify the claim $\|P_\nu\psi_t\|^2 \approx \dd_\nu/\D$.

The basic (trivial) observation about normality is this:

\begin{lem}
For arbitrary $\Hilbert, H, \decomp,\psi_0$ with $\|\psi_0\|=1$ and any $\varepsilon>0$ and $\delta'>0$, if
\be\label{cond3}
G=G(H,\decomp,\psi_0,\nu):=\overline{\Bigl|\|P_\nu\psi_t\|^2-\frac{\dd_\nu}{\D}\Bigr|^2}
< \varepsilon^2 \frac{\dd_\nu}{N\D}\frac{\delta'}{N} =: \mathrm{bound}_1
\ee
for every $\nu$ then the system is $\varepsilon$-$\delta'$-normal in von Neumann's sense. If
\be\label{cond1}
G
< \varepsilon^2 \frac{\dd_\nu^2}{\D^2}\frac{\delta'}{N}=:\mathrm{bound}_2
\ee
for every $\nu$ then the system is $\varepsilon$-$\delta'$-normal in the strong sense.
\end{lem}

\proof
If a non-negative quantity $f(t)$ (such as the $|\cdots|^2$ above) is greater than or equal to $a:=\varepsilon^2 \dd_\nu/N\D>0$ for more than the fraction $b:=\delta'/N>0$ of the time interval $[0,T]$ then its average over $[0,T]$ must be greater than $ab$. By assumption \eqref{cond3}, this is not the case for any $\nu$ when $T$ is sufficiently large. But $|\cdots|^2\geq a$ means violating \eqref{vNdef}. Therefore, for sufficiently large $T$, the fraction of the time when \eqref{vNdef} is violated for any $\nu$ is no greater than $\delta'$; thus, \eqref{mostt} holds with $p(t)$ given by $\forall \nu$ : \eqref{vNdef}. 

In the same way one obtains \eqref{strongdef} from \eqref{cond1}.
\endproof

\section{Von Neumann's QET}
\label{sec:vN}

We now describe von Neumann's result. To evaluate the expression $G$, let $\phi_1,\ldots,\phi_\D$ be an orthonormal basis of $\Hilbert$ consisting of eigenvectors of the Hamiltonian $H$ with eigenvalues $E_1,\ldots,E_\D$, and expand $\psi_0$ in that basis:
\be
\psi_0 = \sum_{\alpha=1}^\D c_\alpha \, \phi_\alpha\,,\quad
\psi_t = \sum_{\alpha=1}^\D e^{-iE_\alpha t} c_\alpha\, \phi_\alpha\,.
\ee
Inserting this into $G$ and multiplying out the square, one obtains
\begin{align}
G
&=\sum_{\alpha,\alpha',\beta,\beta'}\overline{e^{i(E_\alpha-E_{\alpha'}-E_\beta+E_{\beta'}) t}} c^*_\alpha c_{\alpha'} c_\beta c^*_{\beta'}
\scp{\phi_\alpha}{P_\nu|\phi_\beta} \scp{\phi_{\alpha'}}{P_\nu|\phi_{\beta'}}^* \nonumber\\
&-2\frac{\dd_\nu}{\D}\Re\sum_{\alpha,\beta}\overline{e^{i(E_\alpha-E_\beta) t}} c^*_\alpha c_\beta\scp{\phi_\alpha}{P_\nu|\phi_\beta}+\frac{\dd_\nu^2}{\D^2}\,.
\label{expression1}
\end{align}
If $H$ is non-degenerate then $E_\alpha-E_\beta$ vanishes only for $\alpha=\beta$, so the time averaged exponential in the last line is $\delta_{\alpha\beta}$. Furthermore, if $H$ has no resonances then the time averaged exponential in the first line of \eqref{expression1} becomes $\delta_{\alpha\alpha'}\delta_{\beta\beta'} + \delta_{\alpha\beta} \delta_{\alpha'\beta'} - \delta_{\alpha\alpha'}\delta_{\beta\beta'}\delta_{\alpha\beta}$, and we have that
\begin{align}
G
&=\sum_{\alpha,\beta} |c_\alpha|^2 |c_\beta|^2
\biggl( \bigl|\scp{\phi_\alpha}{P_\nu|\phi_\beta}\bigr|^2 + \scp{\phi_\alpha}{P_\nu|\phi_\alpha} \scp{\phi_\beta}{P_\nu|\phi_\beta} \biggr) \nonumber\\
&\quad-\sum_\alpha |c_\alpha|^4 \scp{\phi_\alpha}{P_\nu|\phi_\alpha}^2
-2\frac{\dd_\nu}{\D}\sum_{\alpha} |c_\alpha|^2 \scp{\phi_\alpha}{P_\nu|\phi_\alpha}+\frac{\dd_\nu^2}{\D^2}\\
&=\sum_{\alpha\neq \beta} |c_\alpha|^2 |c_\beta|^2
\bigl|\scp{\phi_\alpha}{P_\nu|\phi_\beta}\bigr|^2 
+\biggl(\sum_\alpha |c_\alpha|^2 
\scp{\phi_\alpha}{P_\nu|\phi_\alpha}-\frac{\dd_\nu}{\D}\bigg)^2\\
&\leq \max_{\alpha\neq \beta}
\bigl|\scp{\phi_\alpha}{P_\nu|\phi_\beta}\bigr|^2 
+ \max_\alpha 
\Bigl(\scp{\phi_\alpha}{P_\nu|\phi_\alpha}-\frac{\dd_\nu}{\D}\Bigr)^2
\end{align}
using $\sum|c_\alpha|^2=1$. This calculation proves the following.

\begin{lem}\label{lem:cond2}
For arbitrary $\Hilbert$ and $\decomp$, for any $H$ without degeneracies and resonances, and for any $\varepsilon>0$ and $\delta'>0$, if, for every $\nu$,
\be\label{cond2}
\max_{\alpha\neq \beta}
\bigl|\scp{\phi_\alpha}{P_\nu|\phi_\beta}\bigr|^2 
+ \max_\alpha 
\Bigl(\scp{\phi_\alpha}{P_\nu|\phi_\alpha}-\frac{\dd_\nu}{\D}\Bigr)^2
< \mathrm{bound}_{1,2}
\ee
then, for every $\psi_0\in\Hilbert$ with $\|\psi_0\|=1$, the system is $\varepsilon$-$\delta'$-normal in von Neumann's sense respectively in the strong sense.
\end{lem}

Note that \emph{every} initial wave function behaves normally, provided $H$ and $\decomp$ together satisfy the condition \eqref{cond2}. Now von Neumann's QET asserts that for any given $H$ and any suitable given values of the $\dd_\nu$, most $\decomp$ will satisfy \eqref{cond2}. It is convenient to think of $\decomp$ as arising from a uniformly distributed orthonormal basis $\omega_1,\ldots,\omega_\D$ in the sense that $\Hilbert_\nu$ is spanned by those $\omega_\alpha$ with $\alpha\in J_\nu$, as described in Section~\ref{sec:most}. The coefficients $U_{\alpha\beta}=\scp{\phi_\beta}{\omega_\alpha}$ of $\omega_\alpha$ relative to the eigenbasis of $H$ then form a Haar-distributed unitary matrix, and
\be
\scp{\phi_\alpha}{P_\nu|\phi_\beta}=\sum_{\gamma\in J_\nu} \scp{\phi_\alpha}{\omega_\gamma}\scp{\omega_\gamma}{\phi_\beta} = \sum_{\gamma\in J_\nu} U_{\gamma\alpha} (U_{\gamma\beta})^*\,.
\ee
Let $\log$ denote the natural logarithm. 

\begin{lem}\label{lem:UvN}
(von Neumann 1929)
There is a (big) constant $C_1>1$ such that whenever the two natural numbers $\D$ and $\dd_\nu$ satisfy
\be
C_1\log \D < \dd_\nu< \frac{\D}{C_1}\,,
\ee
and $U$ is a Haar-distributed random unitary $\D\times\D$ matrix, then
\be
\EEE \max_{\alpha \neq \beta=1}^\D \Bigl| \sum_{\gamma=1}^{\dd_\nu} U_{\gamma\alpha} (U_{\gamma\beta})^*\Bigr|^2 \leq \frac{\log \D}{\D}\,,
\ee
\be
\EEE \max_{\alpha=1}^\D \Bigl(\sum_{\gamma=1}^{\dd_\nu} |U_{\gamma\alpha}|^2 -\frac{\dd_\nu}{\D}\Bigr)^2 \leq \frac{9\dd_\nu \log \D}{\D^2}\,.
\ee
\end{lem}

To express that $\mu\{x|p(x)\}\ge 1-\delta$, we also say that $p(x)$ \emph{holds for $(1-\delta)$-most $x$}. Putting together Lemma~\ref{lem:cond2} (for bound$_1$) and Lemma~\ref{lem:UvN}, we have the following:\footnote{For clarity we have modified von Neumann's statement a bit.}

\begin{thm}\label{thm:vN}
(von Neumann's QET, 1929)
Let $\varepsilon>0$, $\delta>0$, and $\delta'>0$. Suppose the numbers $\D$, $N$, and $\dd_1,\ldots,\dd_N$ are such that $\dd_1+ \ldots +\dd_N=\D$ and, for all $\nu$,
\be\label{cond5}
\max\Bigl(C_1, \frac{10N^2}{\varepsilon^2\delta'\delta} \Bigr) \log \D < \dd_\nu < \D/C_1\,,
\ee
where $C_1$ is the constant of Lemma~\ref{lem:UvN}. For arbitrary $\Hilbert$ of dimension $\D$ and any $H$ without degeneracies and resonances, 
$(1-\delta)$-most orthogonal decompositions $\decomp=\{\Hilbert_\nu\}$ of $\Hilbert$ with $\dim\Hilbert_\nu = \dd_\nu$ are such that for every wave function $\psi_0\in\Hilbert$ with $\|\psi_0\|=1$ the system is $\varepsilon$-$\delta'$-normal in von Neumann's sense. 
\end{thm}

\proof
Regard $\decomp$ as random with uniform distribution and let $X$ be the left hand side of \eqref{cond2}. Using \eqref{cond5}, it follows from Lemma~\ref{lem:UvN} that $\EEE X\leq 10 \log \D/\D$. By Markov's inequality,
\be
\PPP(X \geq \mathrm{bound}_1) \leq \frac{\EEE X}{\mathrm{bound}_1} 
\leq \frac{10\log \D}{\D \,\mathrm{bound}_1}< \delta\,,
\ee
using \eqref{cond5} again. Theorem~\ref{thm:vN} then follows from Lemma~\ref{lem:cond2}.
\endproof

\section{Strong Version}
\label{sec:strong}

It is an unsatisfactory feature of the QET that all $\dd_\nu$ are assumed to be much smaller (by at least a factor $C_1$) than $\D$, an assumption excluding that one of the macro-states $\nu$ corresponds to thermal equilibrium. However, this assumption can be removed, and even the strong sense of normality can be concluded. An inspection of von Neumann's proof of Lemma~\ref{lem:UvN} reveals that it actually proves the following.

\begin{lem}\label{lem:better}
(von Neumann 1929)
There is a (big) constant $C_2>1$ such that whenever the two natural numbers $\D$ and $\dd_\nu$ satisfy
\be
C_2<\dd_\nu<\D-C_2\,,
\ee
and $U$ is a Haar-distributed random unitary $\D\times\D$ matrix then, for every $0<a<\dd_\nu^2/\D^2 C_2$,
\be\label{estimate1}
\PPP\biggl( \max_{\alpha \neq \beta=1}^\D \Bigl| \sum_{\gamma=1}^{\dd_\nu} U_{\gamma\alpha} (U_{\gamma\beta})^*\biggr|^2 \geq a \Bigr) \leq
\frac{\D^2}{2} \exp\Bigl( -4a(\D-1) \Bigr)\,,
\ee
\be\label{estimate2}
\PPP\biggl( \max_{\alpha=1}^\D \Bigl(\sum_{\gamma=1}^{\dd_\nu} |U_{\gamma\alpha}|^2 -\frac{\dd_\nu}{\D}\Bigr)^2\geq a \biggr) \leq
\frac{\D^3}{\sqrt{2\pi \dd_\nu} (\D-\dd_\nu)} \exp\Bigl( -\Theta \frac{\D^2 a}{2\dd_\nu} \Bigr)\,.
\ee
with $\Theta = 1-\frac{2}{3\sqrt{C_2}}$.
\end{lem}

From this we can obtain, with Lemma~\ref{lem:cond2}, the following stronger version of the QET, which von Neumann did not mention.

\begin{thm}\label{thm:strong}
Theorem~\ref{thm:vN} remains valid if one replaces ``normal in von Neumann's sense'' by ``normal in the strong sense'' and \eqref{cond5} by
\be\label{cond6}
\max\Bigl(C_2, \sqrt{(3N/\varepsilon^2\delta') \D \log \D}\Bigr) <
\dd_\nu < \D-C_2\,,
\ee
\be\label{cond7}
\varepsilon^2\delta'<2N/C_2\,, \quad
\D/\log \D> 100N/\varepsilon^2\delta'\,, \quad\text{and}\quad
\D>1/\delta\,,
\ee
where $C_2$ is the constant of Lemma~\ref{lem:better}.
\end{thm}

\proof
Set $a=\mathrm{bound}_2/2=(\varepsilon^2\delta'/2N)(\dd_\nu/\D)^2$ in \eqref{estimate1} and \eqref{estimate2}. The first assumption in \eqref{cond7} ensures that the condition $a<\dd_\nu^2/\D^2 C_2$ in Lemma~\ref{lem:better} is satisfied. The assumption \eqref{cond6} includes
\begin{align}
\dd_\nu^2 &> (3N/\varepsilon^2\delta') \D \log \D \label{cond61squared}\\
&> (N/\varepsilon^2\delta') \D (2\log \D-\log \delta)
\label{ddnu2est}
\end{align}
using $\log \D > -\log \delta$ from the third assumption in \eqref{cond7}. Now \eqref{ddnu2est} implies that $4a(\D-1)>2a\D\geq 2\log \D-\log \delta$, so that the right hand side of \eqref{estimate1} is less than $\delta/2$. Furthermore, from the second assumption in \eqref{cond7} 
we have that $1>100 N \log \D/\varepsilon^2 \delta' \D$, which 
yields with \eqref{cond61squared} that $\dd_\nu^2 > (300N^2/\varepsilon^4 \delta^{\prime2}) \log^2 \D$, and thus $\dd_\nu > (16N/\Theta \varepsilon^2 \delta') \log \D$, using $\Theta>16/\sqrt{300}$ (which follows from $C_2\geq 121$). Because of $\log\D>-\log \delta$, we have that
\be
\dd_\nu> (4N/\Theta \varepsilon^2\delta') (3\log \D-\log \delta)\,,
\ee
which implies that $\Theta \D^2 a/2\dd_\nu = \Theta (\varepsilon^2\delta'/4N)\dd_\nu > 3\log \D-\log \delta$, so also the right hand side of \eqref{estimate2} is less than $\delta/2$. Thus, \eqref{cond2} is fulfilled for bound$_2$ with probability at least $1-\delta$.
\endproof

The stronger conclusion requires the strong assumption that $\sqrt{\D\log \D}\ll \dd_\nu$ whereas von Neumann's version needed $\log \D \ll \dd_\nu \ll \D$. 

\z{Concerning a thermal equilibrium macro-state with $\dd_{eq}/\D \geq 1-\varepsilon$, Theorem~\ref{thm:strong} provides conditions under which most subspaces $\Hilbert_{eq}$ of dimension $\dd_{eq}$ are such that, for every $\psi_0\in\Hilbert$ with $\|\psi_0\|=1$, the system will be in thermal equilibrium for most times. More precisely, Theorem~\ref{thm:strong} implies the following: 
\textit{Let $\varepsilon>0$, $\delta>0$, and $\delta'>0$. Suppose that the number $\D$ is so big that \eqref{cond7} holds with $N=2$, and that $\dd_{eq}$ is such that 
\be
1-\varepsilon \leq \frac{\dd_{eq}}{\D} \leq 1\,,
\ee
\be\label{cond8}
\max\Bigl(C_2, \sqrt{(6/\varepsilon^2\delta') \D \log \D}\Bigr) <
\dd_{eq} < \D-
\max\Bigl(C_2, \sqrt{(6/\varepsilon^2\delta') \D \log \D}\Bigr)\,.
\ee
For arbitrary $\Hilbert$ of dimension $\D$ and any Hamiltonian $H$ without degeneracies and resonances, 
$(1-\delta)$-most subspaces $\Hilbert_{eq}$ of $\Hilbert$ with $\dim\Hilbert_{eq} = \dd_{eq}$ are such that for every wave function $\psi_0\in\Hilbert$ with $\|\psi_0\|=1$, the relation
\be
\|P_{eq}\psi_t\|^2 > 1-2\varepsilon
\ee
holds for $(1-\delta')$-most $t$.}
In this statement, however, the conditions can be relaxed (in particular, $H$ may have resonances, and the upper bound on $\dd_{eq}$ in \eqref{cond8} can be replaced with $\D$), and the statement can be obtained through a proof that is much simpler than von Neumann's; see \cite{GLMTZ09}.}

\section{Misrepresentations}
\label{sec:overview}

We now show that the statements presented as the QET in \cite{FL57,BL58} differ from the original theorem (in fact in inequivalent ways) and are dynamically vacuous.

It is helpful to introduce the symbol $\formost$ to denote ``for most.'' It can be regarded as a quantifier like the standard symbols $\forall$ (for all) and $\exists$ (for at least one). So, if $p(x)$ is a statement containing the free variable $x$ then we write $\formost x: \:p(x)$ when we mean $\mu\{x|p(x)\}\geq 1-\delta$, assuming that it is clear from the context which measure $\mu$ and which magnitude of $\delta$ are intended. With this notation, the misunderstanding as described in \eqref{allpsimostDp} versus \eqref{mostDallpsip} can be expressed by saying that the quantifiers $\formost x$ and $\forall y$ do not commute:
\be\label{noncommute}
\formost x \forall y:p(x,y) 
\quad \not\Leftrightarrow \quad 
\forall y \formost x:p(x,y)\,. 
\ee
The two expressions are not equivalent. Indeed, the set of $x$'s (whose measure is close to 1) is allowed to depend on $y$ if the quantifiers are of the form $\forall y \formost x$ but not if they are of the form $\formost x \forall y$. That is, if they are \x{of the form} $\formost x \forall y$ then there exists a set $M$ of $x$'s, not depending on $y$, with $\mu_x(M)\geq 1-\delta$ such that $\forall x\in M\forall y: p(x,y)$. Thus the first expression in \eqref{noncommute} is stronger than the second:
\be\label{imp}
\formost x \forall y:p(x,y) 
\quad \Rightarrow \quad 
\forall y \formost x:p(x,y)\,. 
\ee

This should be contrasted with situations in which quantifiers do commute, for example $\forall x \forall y \Leftrightarrow \forall y \forall x$ and $\formost x \formost y \Leftrightarrow \formost y \formost x$ (though the bound $\delta$ on the exceptions may become worse\footnote{More precisely, if
\be\label{ineq1}
\mu_x\bigl\{x\big| \mu_y\{y|p(x,y)\}\geq 1-\delta_y \bigr\} \geq 1-\delta_x
\ee
then, for every $\varepsilon_x>0$,
\be\label{ineq2}
\mu_y\bigl\{ y\big| \mu_x\{x|p(x,y)\}\geq 1-\varepsilon_x \bigr\}\geq 1-\varepsilon_y
\ee
with $\varepsilon_y\geq (\delta_x+\delta_y-\delta_x\delta_y)/\varepsilon_x$. (For example, \eqref{ineq2} holds for $\varepsilon_x=\varepsilon_y=\sqrt{\delta_x+\delta_y}$.) To see this, note that \eqref{ineq1} implies that, relative to the product measure $\mu_x \otimes \mu_y$, at least the fraction $(1-\delta_x)(1-\delta_y)$ of all pairs $(x,y)$ satisfies $p(x,y)$; thus, 
\[
\int \mu_y(dy) \, \mu_x\{x|p(x,y)\} = \mu_x\otimes \mu_y \{(x,y)|p(x,y)\} \geq 1-(\delta_x+\delta_y-\delta_x\delta_y)\,,
\]
and this implies \eqref{ineq2}.}). An exceptional case, in which $\formost x$ and $\forall y$ do commute, occurs when the variable $y$ assumes only a very limited number $n$ (e.g., $n=10$) of possible values: Then $\forall y \formost x:p$ implies $\formost x \forall y:p$ with, however, the bound $\delta$ on the exceptions worse by a factor of $n$, $\delta\to n\delta$. In our case, however, $y=\psi_0$ varies in an infinite set.

In this symbolic notation, and leaving out some details, Theorems~\ref{thm:vN} and \ref{thm:strong} can be paraphrased as:
\be\label{vNsummary}
\forall H \:\formost \decomp \:\forall \psi_0 \:\formost t \:\forall \nu: \|P_\nu\psi_t\|^2 \approx \dd_\nu/\D\,,
\ee
where $\forall H$ should be taken to mean ``for all Hamiltonians without degeneracies and resonances,'' and $\approx$ should be understood either in the wide sense of \eqref{vNdef} for Theorem~\ref{thm:vN}, or in the sense of \eqref{strongdef} for Theorem~\ref{thm:strong}. Let us now look at what \cite{FL57,BL58} write.

\bigskip

We focus first on the article of Bocchieri and Loinger \cite{BL58}. As we show presently, their version of the QET has a different order of quantifiers, with fatal consequences. It also differs in a second way from the original as it deals with the strong sense of normality instead of von Neumann's sense; this, of course, is a strengthening of von Neumann's statement. Finally, their version drops von Neumann's hypotheses on the Hamiltonian (no degeneracy, no resonance); this, of course, is a difference that Bocchieri and Loinger were aware of and emphasized as evidence that von Neumann made unnecessary hypotheses. 

Indeed, in \cite{BL58}, the statement  ``These relations constitute von Neumann's ergodic theorem'' (p.~670) is preceded by their Eq.~(13), which in our notation reads
\be\label{BL58}
\EEE \overline{\|P_\nu\psi_t\|^2} = \frac{\dd_\nu}{\D}\,; \quad
\frac{\EEE \overline{\bigl|\|P_\nu\psi_t\|^2-\dd_\nu/\D\bigr|^2}}{\dd_\nu^2/\D^2} \ll 1\,,
\ee
where the average $\EEE$ is taken over $\decomp$ relative to the uniform distribution.\footnote{More precisely, their proof shows that for every $\eta>0$ and every $H$, if every $\dd_\nu>1/\eta$ then, for all $\psi_0$ and $\nu$,
$\EEE \overline{\bigl|\|P_\nu\psi_t\|^2-\dd_\nu/\D\bigr|^2} < \eta \dd_\nu^2/\D^2$.} 
From this it follows that for all $\psi_0$ it is true for most $\decomp$ that $\|P_\nu \psi_t\|^2 \approx \dd_\nu/\D$ for most $t$, with deviation small compared to $\dd_\nu/\D$. Moreover, as \eqref{BL58} holds for all $H$, and, via \eqref{cond1}, the conclusion can be shown to hold simultaneously for all $\nu$, the version of \cite{BL58} can be written, in analogy to \eqref{vNsummary}, as
\be\label{BL58summary}
\forall H\: \forall \psi_0 \:\formost\decomp \:\formost t \:\forall \nu: \|P_\nu\psi_t\|^2 \approx \dd_\nu/\D\,.
\ee

This statement is not only inequivalent to von Neumann's, it is also \emph{dynamically vacuous}. By this we mean that it follows from a statement that does not refer to any time other than 0. Indeed, the relations \eqref{BL58} are proved in \cite{BL58} by first proving for any fixed $\psi$ that\footnote{In fact, these expectation values are independent of $\psi$, by the invariance of the Haar measure.}
\be\label{BL1}
\EEE \|P_\nu\psi\|^2 = \frac{\dd_\nu}{\D}\,; \quad
\frac{\EEE \bigl|\|P_\nu\psi\|^2-\dd_\nu/\D\bigr|^2}{\dd_\nu^2/\D^2} \ll 1\,,
\ee
which is \eqref{BL58} without the procedure of time averaging,
then setting $\psi=\psi_t$ and taking the time average on both relations, and finally commuting the time average and the average $\EEE$ over $\decomp$, which is always allowed by Fubini's theorem. In the notation using the symbol $\formost$, \eqref{BL1} yields
\be\label{BL1summary}
\forall \psi \:\formost\decomp \:\forall \nu: \|P_\nu\psi\|^2 \approx \dd_\nu/\D\,.
\ee
This fact is the non-dynamical reason why \eqref{BL58summary} is true: Since \eqref{BL1summary} applies to every $\psi$, it applies in particular to $\psi_t$ for any $H$, $\psi_0$, and $t$. That is, \eqref{BL1summary} implies
\be\label{BL2}
\forall H\: \forall \psi_0 \:\forall t \:\formost\decomp \:\forall \nu: \|P_\nu\psi_t\|^2 \approx \dd_\nu/\D\,,
\ee
and since $\forall t \:\formost \decomp \Rightarrow \formost t\: \formost \decomp \Rightarrow \formost \decomp \:\formost t$, \eqref{BL2} implies \eqref{BL58summary}. Thus, \eqref{BL58summary} is dynamically vacuous. This fact was essentially the criticism put forward against the QET in \cite{BL58}.\footnote{The exact nature of the criticism, though, remained a bit unclear in \cite{BL58}, as Bocchieri and Loinger did not make explicit what it means for a statement to be dynamically vacuous. They pointed out that \eqref{BL58} is valid for every Hamiltonian, including $H=0$, and that the proof of \eqref{BL58} by means of \eqref{BL1} did not, in fact, require that $\psi_t=\exp(-iHt)\psi_0$, but only that $\psi_t=f_t(\psi_0)$ for an arbitrary measure-preserving mapping $f_t$ from the unit sphere to itself. These facts strongly suggest that \eqref{BL58} is dynamically vacuous, but should per se not be regarded as a proof; for example, the Poincar\'e recurrence theorem \cite{recurrence} is valid for every Hamiltonian, or in fact for every measure-preserving flow $f_t$ on the unit sphere in a finite-dimensional Hilbert space, but clearly has dynamical content. That is why we defined a ``dynamically vacuous statement'' to be a logical consequence of a statement that does not refer to time.}

\bigskip

We turn to the article of Farquhar and Landsberg \cite{FL57}. As we show presently, their version of the QET differs from the original in the same ways as the version of \cite{BL58}, as well as in that it concerns only the \emph{time average} of $\|P_\nu\psi_t\|^2$, while the original QET concerns the value of $\|P_\nu\psi_t\|^2$ \emph{for most $t$}.

Indeed, the result on which their version of the QET is based is expressed in their Eq.~(2.17), which holds for every $H$ and $\D\geq 3$ and reads in our notation as
\be\label{FL57}
\frac{\EEE \bigl|\overline{\|P_\nu\psi_t\|^2}-\dd_\nu/\D \bigr|^2}{\dd_\nu^2/\D^2}
<\frac{2(\D-\dd_\nu)}{\dd_\nu\D}\,.
\ee
For large $\dd_\nu$, this yields
\be\label{FL3}
\frac{\EEE \bigl|\overline{\|P_\nu\psi_t\|^2}-\dd_\nu/\D \bigr|^2}{\dd_\nu^2/\D^2}
\ll 1\,,
\ee
and thus
\be\label{FL57summary}
\forall H\: \forall \psi_0 \:\formost\decomp \:\forall \nu: \overline{\|P_\nu\psi_t\|^2} \approx \dd_\nu/\D\,.
\ee
This result concerns only the time average of $\|P_\nu\psi_t\|^2$ but provides no control over the time variance, and so does not inform us about the behavior for \emph{most} $t$. Moreover, \eqref{FL57summary} has the wrong order of quantifiers. Finally, since \eqref{FL3} follows from the inequality in \eqref{BL58} using $\overline{f(t)}^2 \leq \overline{f(t)^2}$, it is a logical consequence of a dynamically vacuous statement, and thus is itself dynamically vacuous.

\section{Typical Hamiltonian}
\label{sec:typicalH}
\newcommand{\A}{\mathscr{A}}
\newcommand{\Tr}{\text{Tr}}

Normality for most $\decomp$s is more or less equivalent to normality for most $H$s. Indeed, by the ``unitary inversion trick'' described in Section~\ref{sec:most}, one can trade the typicality assumption on $\decomp$ in the QET for a typicality assumption on $H$, without any essential modification of the proof. This is because the relevant condition \eqref{cond2} involves only
\be
\scp{\phi_\alpha}{P_\nu|\phi_\beta}=\sum_{\gamma\in J_\nu} \scp{\phi_\alpha}{\omega_\gamma}\scp{\omega_\gamma}{\phi_\beta}\,,
\ee
where we can either regard $\phi$ as fixed and $\omega$ as random (as von Neumann did) or vice versa. With this change, the (strong) QET reads as follows.

\begin{thm}\label{thm:typH}
Let $\varepsilon>0$, $\delta'>0$, and $\delta>0$. Suppose the numbers $\D$, $N$, and $\dd_1+\ldots+\dd_N=\D$ satisfy \eqref{cond6} and \eqref{cond7}. Suppose further that the real numbers $E_1,\ldots,E_\D$ are all distinct and have no resonances as defined in \eqref{noresonance}. For arbitrary $\Hilbert$ of dimension $\D$ and any orthogonal decomposition $\decomp=\{\Hilbert_\nu\}$ with $\dim\Hilbert_\nu = \dd_\nu$, $(1-\delta)$-most operators $H$ with eigenvalues $E_1,\ldots, E_\D$ are such that for every wave function $\psi_0\in\Hilbert$ with $\|\psi_0\|=1$ the system is $\varepsilon$-$\delta'$-normal in the strong sense. 
\end{thm}

This means, in the notation of \eqref{vNsummary}, that
\be\label{Hsummary}
\forall \decomp \:\formost H \:\forall \psi_0 \:\formost t \:\forall \nu: \|P_\nu\psi_t\|^2 \approx \dd_\nu/\D\,.
\ee
It would be nice also to have a similar theorem asserting that normality for all $\psi_0$ is typical even within a smaller class of Hamiltonians, say those of the form
\be
H=-\sum_{i=1}^n \frac{\hbar^2\nabla_i^2}{2m_i} + \sum_{i=1}^n U(x_i) + 
\sum_{i\neq j}V(x_i-x_j)\,,
\ee
where the pair potential $V$ is allowed to be any function from a suitable class. Here, $n$ denotes the number of particles, $x_i\in\RRR^3$ the coordinate of particle $i$, $\nabla_i$ the derivative relative to $x_i$, $m_i$ the mass of particle $i$, and $U$ the external potential. However, such a theorem seems presently out of reach.



As a corollary of \eqref{Hsummary}, one obtains for $\nu=eq$ that
\be\label{eqsummary}
\forall \Hilbert_{eq} \:\formost H \:\forall \psi_0 \:\formost t : \|P_{eq}\psi_t\|^2 \approx 1\,,
\ee
where $\forall \Hilbert_{eq}$ should be taken to mean ``for all subspaces $\Hilbert_{eq}$ of dimension $\dd_{eq}$'' (which is greater than $(1-\varepsilon')\D$). 
In fact, this conclusion remains true \cite{GLMTZ09} under weaker technical assumptions ($H$ may have resonances, and \eqref{cond6} can be replaced by $(1-\varepsilon')\D<\dd_{eq}\leq \D$).

As a corollary of \eqref{eqsummary}, for a typical Hamiltonian every energy eigenfunction  is in thermal equilibrium, i.e., close to $\Hilbert_{eq}$. (This statement could, of course, be obtained more directly: The condition that every energy eigenfunction is in equilibrium is a special case, for $\nu=eq$, of the condition $\scp{\phi_\alpha}{P_\nu|\phi_\alpha}\approx \dd_\nu/\D$ for all $\alpha$, which is part of condition \eqref{cond2}, which by Lemma~\ref{lem:UvN} is typically obeyed.)

We can be a bit more general than either Theorem~\ref{thm:strong} or Theorem~\ref{thm:typH} and say that what is needed to obtain strong normality is that the unitary matrix $U_{\alpha\beta}=\scp{\phi_\beta}{\omega_\alpha}$ relating the energy eigenbasis $\phi_\beta$ to a basis $\omega_\alpha$ aligned with $\decomp$ be like most unitary matrices in that they satisfy \eqref{cond2}. This means, more or less, that the energy eigenbasis and $\decomp$ should be unrelated. By the way, this is connected to the reason why $\Hilbert$ was physically interpreted as a ``micro-canonical'' space, i.e., one corresponding to an ``energy shell'': For a more comprehensive Hilbert space including states of macroscopically different energies, the energy eigenbasis and $\decomp$ would no longer be unrelated. Indeed, a sufficiently coarse-grained version of the Hamiltonian should be among the macroscopic observables and thus be diagonal in the $\omega_\alpha$ basis.

\section{Comparison with Recent Literature}
\label{sec:history}

The results of \cite{T98,R08,LPSW08} also concern conditions under which a quantum system will spend most of the time in ``thermal equilibrium.'' For the sake of comparison, their results, as well as von Neumann's, can be described in a unified way as follows. Let us say that a system with initial wave function $\psi(0)$ \emph{equilibrates} relative to a class $\A$ of observables if for most times $\tau$,
\be\label{equidef}
\scp{\psi(\tau)}{A|\psi(\tau)} \approx 
\Tr\Bigl(\overline{\ket{\psi(t)}\bra{\psi(t)}}A\Bigr) 
\text{ for all }A\in\A\,.
\ee
We then say that the system \emph{thermalizes} relative to $\A$ if it equilibrates and, moreover,
\be
\Tr\Bigl(\overline{\ket{\psi(t)}\bra{\psi(t)}} A\Bigr)\approx
\Tr\bigl(\rho_{mc}A\bigr) \text{ for all }A\in\A\,,
\ee
with $\rho_{mc}$ the micro-canonical density matrix (in our notation, $1/\D$ times the projection $P$ to $\Hilbert$). With these definitions, the results of \cite{T98,R08,LPSW08} can be formulated by saying that, under suitable hypotheses on $H$ and $\psi(0)$ and for large enough $\D$, a system will equilibrate, or even thermalize, relative to a suitable class $\A$.
Von Neumann's quantum ergodic theorem establishes thermalization for a family $\A$ of commuting observables,  the algebra generated by $\{M_1,\ldots,M_k\}$ in the notation of Section~\ref{introduction}. 

Tasaki \cite{T98} as well as Linden, Popescu, Short, and Winter \cite{LPSW08} consider a system coupled to a heat bath, $\Hilbert_\mathrm{total}=\Hilbert_\mathrm{sys}\otimes\Hilbert_\mathrm{bath}$, and take $\A$ to contain all operators of the form $A_\mathrm{sys}\otimes 1_\mathrm{bath}$. Tasaki considers a rather special class of Hamiltonians and establishes thermalization assuming that 
\be
\max_\alpha |\scp{\phi_\alpha}{\psi(0)}|^2 \ll 1\,,
\ee
a condition that implies that many eigenstates of $H$ contribute to $\psi(0)$ appreciably and that can (more or less) equivalently be rewritten as
\be\label{contribute}
\sum_\alpha \bigl|\scp{\phi_\alpha}{\psi(0)}\bigr|^4 \ll 1\,.
\ee
Under the assumption \eqref{contribute} on $\psi(0)$,
Linden et al.\ establish equilibration for $H$ satisfying \eqref{noresonance}. They also establish a result in the direction of thermalization under the additional hypothesis that the dimension of the energy shell of the bath is much greater than $\dim \Hilbert_\mathrm{sys}$.

Reimann's mathematical result \cite{R08} can be described in the above scheme as follows. Let $\A$ be the set of all observables $A$ with (possibly degenerate) eigenvalues between 0 and 1 such that the absolute difference between any two eigenvalues is at least (say) $10^{-1000}$. He establishes equilibration for $H$ satisfying \eqref{noresonance}, assuming that $\psi(0)$ satisfies \eqref{contribute}.

\bigskip

\noindent\textit{Acknowledgements.} 
We thank Detlef D\"urr  and Tony Short for helpful discussions. 
S.~Goldstein was supported in part by National Science Foundation [grant DMS-0504504].
N.~Zangh\`\i\ is supported in part by Istituto Nazionale di Fisica Nucleare. 
J.~L.~Lebowitz is supported in part by NSF [grant DMR 08-02120] and by AFOSR [grant AF-FA 09550-07].


\begin{thebibliography}{14}


\bibitem{BL58} P. Bocchieri, A. Loinger:
                    Ergodic Theorem in Quantum Mechanics.
                    \textit{Physical Review} \textbf{111}: 668--670 (1958)

\bibitem{BL59} P. Bocchieri, A. Loinger:
                    Ergodic Foundation of Quantum Statistical Mechanics.
                    \textit{Physical Review} \textbf{114}: 948--951 (1959)


\bibitem{Choi88} 
	M. D. Choi: Almost commuting matrices need not be nearly commuting.
	\textit{Proc. Amer. Math. Soc.} \textbf{102(3)}: 529--533 (1988)

\bibitem{DRMN06} W. De Roeck, C. Maes and  K. Neto\v{c}n\'y:  
	Quantum Macrostates, Equivalence of Ensembles and an $H$-theorem.
	\textit{J. Math. Phys.} \textbf{47}: 073303 (2006).

\bibitem{F61} I. E. Farquhar:
	The Present State of Ergodic Theory.
	\textit{Nature} \textbf{190}: 17--18 (1961)

\bibitem{Fbook} I. E. Farquhar: 
	\textit{Ergodic Theory in Statistical Mechanics.}
	Interscience Publishers and John Wiley (1964)

\bibitem{FL57} I. E. Farquhar, P. T. Landsberg:
	On the quantum-statistical ergodic and $H$-theorems.
	\textit{Proc. Royal Soc. London A} \textbf{239}: 134--144 (1957)

\bibitem{F55} M. Fierz:
	Ergodensatz in der Quantenmechanik.
	\textit{Helvetica Physica Acta} \textbf{28}: 705--715 (1955)

\bibitem{GMM04} J. Gemmer, M. Michel, G. Mahler:
	\textit{Quantum Thermodynamics}. 
	Lecture Notes in Physics \textbf{657}, Springer-Verlag (2004)

\bibitem{GLMTZ09} S. Goldstein, J. L. Lebowitz, C. Mastrodonato, R. Tumulka, N. Zangh\`\i:
	On the Approach to Thermal Equilibrium of Macroscopic Quantum Systems.
	\textit{Phys. Rev. E} \textbf{81}: 011109 (2010). 
	\url{http://arxiv.org/abs/0911.1724}

\bibitem{GLTZ06} S. Goldstein, J. L. Lebowitz, R. Tumulka, N. Zangh\`\i:
                            Canonical Typicality.
                            \textit{Phys. Rev. Lett.} \textbf{96}, 050403 (2006).
                            \url{http://arxiv.org/abs/cond-mat/0511091}

\bibitem{GLTZ10} S. Goldstein, J. L. Lebowitz, R. Tumulka, N. Zangh\`\i:
	Long-Time Behavior of Macroscopic Quantum Systems: Commentary 	Accompanying the English Translation of John von Neumann's 1929 
	Article on the Quantum Ergodic Theorem. 
	To appear in \textit{European Phys. J. H} (2010).
	\url{http://arxiv.org/abs/1003.2129}


\bibitem{J63} R. Jancel: 
	\textit{Foundations of Classical and Quantum Statistical Mechanics}.
	Oxford: Pergamon (1969).
	Translation by W. E. Jones of
	\textit{Les Fondements de la M\'ecanique Statistique Classique e Quantique.}
	Paris: Gauthier-Villars (1963)

\bibitem{Lan61} P. T. Landsberg:
	Quantum Statistical Ergodic and $H$-Theorems for Incompletely Specified 
	Systems.
	\textit{Proc. Royal Soc. London A} \textbf{262}: 100--109 (1961) 

\bibitem{Lan05} P. T. Landsberg:
	Pauli, an ergodic theorem and related matters.
	\textit{American Journal of Physics} \textbf{73(2)}: 119--121 (2005)


\bibitem{LPSW08}
	N. Linden, S. Popescu, A. J. Short, A. Winter:
	Quantum mechanical evolution towards thermal equilibrium. 
	\textit{Phys. Rev. E} \textbf{79}: 061103 (2009).
	\url{http://arxiv.org/abs/0812.2385}

\bibitem{Lud58} G. Ludwig:
	Zum Ergodensatz und zum Begriff der makroskopischen Observablen. I.
	\textit{Zeitschrift f\"ur Physik} \textbf{150}: 346--374 (1958)

\bibitem{normalnumber} Normal number. 
	In \emph{Wikipedia, the free encyclopedia} (accessed June 4, 2009), 
	\url{http://en.wikipedia.org/wiki/Normal_number}

\bibitem{PF37} W. Pauli, M. Fierz:
	\"Uber das $H$-Theorem in der Quantenmechanik.
	\textit{Zeitschrift f\"ur Physik} \textbf{106}: 572--587 (1937)

\bibitem{P56} W. Pauli in a letter to M. Fierz, dated 9 August 1956, quoted from \cite{Lan05}.

\bibitem{Pech84} Ph. Pechukas:
	Sharpening an inequality in quantum ergodic theory.
	\textit{J. Math. Phys.} \textbf{25}: 532 (1984)


\bibitem{recurrence} Poincar\'e recurrence theorem. 
	In \emph{Wikipedia, the free encyclopedia} (accessed June 9, 2009), 
	\url{http://en.wikipedia.org/wiki/Poincare_recurrence_theorem}

\bibitem{PSW06} S. Popescu, A. J. Short, A. Winter:
  Entanglement and the foundation of statistical mechanics.
  \textit{Nature Physics} \textbf{21}(11): 754--758 (2006)

\bibitem{PS60} G. M. Prosperi, A. Scotti: 
	Ergodicity Conditions in Quantum Mechanics.
	\textit{J. Math. Phys.} \textbf{1}: 218 (1960)

\bibitem{R08} P. Reimann: 
	Foundation of Statistical Mechanics under Experimentally Realistic Conditions. 
	\textit{Phys. Rev. Lett.} \textbf{101}: 190403 (2008)

\bibitem{RDO08} M. Rigol, V. Dunjko, M. Olshanii:
	Thermalization and its mechanism for generic isolated quantum systems.
	\textit{Nature} \textbf{452}: 854--858 (2008)

\bibitem{schr} E. Schr\"odinger: 
	Energieaustausch nach der Wellenmechanik.
	\textit{Annalen der Physik} \textbf{388(15)}: 956--968 (1927). 
	English translation by J. F. Shearer: 
	The Exchange of Energy according to Wave Mechanics, 
	pp. 137--146 in E. Schr\"odinger: \textit{Collected Papers on Wave Mechanics.} 
	Providence, R.I.: AMS Chelsea (1928 and 1982)


\bibitem{T98} H. Tasaki: 
	From Quantum Dynamics to the Canonical
	Distribution: General Picture and a Rigorous Example. 
	\textit{Phys. Rev. Lett.} \textbf{80}: 1373--1376 (1998)

\bibitem{TKS91} M. Toda, R. Kubo, N. Saito: 
	\textit{Statistical Physics I.} 
	Springer-Verlag (1991)

\bibitem{vH59} L. van Hove: 
	Ergodic  Behaviour of Quantum Many-Body Systems. 
	\textit{Physica} \textbf{25}: 268--276 (1959)


\bibitem{vN29} J. von Neumann:
      Beweis des Ergodensatzes und des $H$-Theorems in der neuen Mechanik. 
      \textit{Zeitschrift f\"ur Physik} \textbf{57}: 30--70 (1929). English translation
      by R. Tumulka in J. von Neumann: Proof of the ergodic theorem and
      the $H$-theorem in quantum mechanics. To appear in 
      \textit{European Phys. J. H} (2010).
      	\url{http://arxiv.org/abs/1003.2133}


\bibitem{Wigner} E. P.  Wigner:  Random Matrices in Physics. \textit{SIAM Review} \textbf{9}: 1--23 (1967)
\end{thebibliography}
\end{document}